\documentclass[12pt,a4paper,showkeys]{revtex4}
\usepackage{graphicx}
\usepackage{epsfig}
\usepackage{bm}
\usepackage{amsmath}
\usepackage{amssymb}
\usepackage{graphics}
\usepackage{epstopdf}
\usepackage[linkcolor=red,citecolor=green,urlcolor=blue]{hyperref}

\newcommand{\pom}{{I\!\!P}}

\newcommand{\xpom}{x_\pom}

\begin{document}

\title{ \textbf{Single-diffractive Drell-Yan pair production at the LHC}}
\author{Federico Alberto Ceccopieri}
\email{federico.alberto.ceccopieri@cern.ch}
\affiliation{Dipartimento di Fisica e Geologia, Universit\`a degli Studi di Perugia, 
INFN, Sezione di Perugia, Italy} 
\affiliation{IFPA, Universit\'e de Li\`ege,  B4000, Li\`ege, Belgium}

\begin{abstract}
\noindent
We present predictions for single-diffractive 
low-mass Drell-Yan pair production in $pp$ collisions at the LHC at $\sqrt{s}=13$ TeV. 
Predictions are obtained adopting a factorised form for the relevant cross sections and are based on a new set of diffractive parton distributions resulting from the QCD analysis of combined HERA leading proton data.  
We discuss a number of observables useful to characterise the expected factorisation breaking effects.
\end{abstract}

\maketitle

\section{Introduction}
\label{intro}

The diffractive physics program pursued at the HERA $ep$ collider 
in the recent past has substantially improved our knowledge on the dynamics of this class of processes. 
In the deep inelastic regime, the presence of a hard scale enables the derivation 
of a dedicated factorisation theorem~\cite{factorisation_soft,factorisation_coll}
which allows the investigation of the partonic structure of the colour singlet exchanged in the $t$-channel.
From scaling violations of the diffractive Deep Inelastic Scattering (DDIS) 
structure functions, quite precise diffractive parton distributions functions (dPDFs) 
have been extracted by performing QCD analysis~\cite{H106LRG,H107dijet,ZEUS09final,myDPDF} of available data.  

With this tool available, factorisation tests have been conducted in order to investigate 
the range of validity of this hypothesis in processes other than DDIS. 
Factorisation has been shown to hold, as expected theoretically, in diffractive dijets production in DIS, 
where NLO predictions based on dPDFs well describe experimental cross sections~\cite{H115dijet,ZEUS09final}
both in shape and normalisation.
Factorisation breaking effects are expected to appear in diffractive photoproduction of dijets 
due to the resolved component of the quasi-real photon.
In such a case, however, H1~\cite{H115dijet} reported a global suppression factor of data over 
NLO theory around 0.5 while ZEUS~\cite{ZEUS08_phpdijet} found the same ratio compatible with unity. 
To date, these conflicting results prevent to draw a conclusive statement about factorisation in this case.
We note, however, that the measurements of diffractive dijet photoproduction in ultraperipheral
collisions in $pp$ and $pA$ collisions at the LHC~\cite{guzey_klasen} may offer an alternative way to settle 
this issue.

Complementary informations on the nature of diffraction has been provided by hard diffraction measurements 
in hadronic collisions. 
As theoretically anticipated in Refs.~\cite{break1,break2,factorisation_soft}
and experimentally observed in $p \bar{p}$ collisions at Tevatron~\cite{diff_bbar,diff_dijet,diff_WZ}, 
factorisation is strongly violated in such a case. 
In particular, predictions based on a factorised expressions for the relevant cross sections in terms of diffractive parton distributions extracted from HERA data overestimate hard diffraction measurements 
by a factor $\mathcal{O}(10)$. This conclusion persists even after the inclusion of higher order QCD corrections~\cite{klasen}.

A rich program at the LHC is being pursued in diffractive physics by all Collaborations 
either based on the identification of large rapidity gaps (LRG)~\cite{LHC_LRG} or by using 
dedicated proton spectrometers~\cite{LHC_rpot}. 
Complementing Tevatron ($\sqrt{s}=1.96$ TeV) results with forthcoming ones from the LHC at higher centre-of-mass energies ($\sqrt{s}=8,13$ TeV) will give information on the energy dependence, if any, of the suppression factor,
the socalled rapidity gap survival (RGS) probability.
Hopefully, they will allow to study its kinematic dependences, among which the one on the scale 
characterising the hard process appears to be particularly relevant. 
In the simplest scenario, it will be possible to clarify
whether factorisation may still hold but revisited in a weak form through a global or local rescaling of diffractive PDFs extracted from DDIS and to study their degree of universality among different hard processes in hadronic collisions.

The purpose of the present paper is to present predictions for the single-diffractive Drell-Yan pair production at the LHC at $\sqrt{s}=13$ TeV, one of the clean and simple measurable process in hadronic collisions.
In such a process, the invariant mass of the lepton pair can be easily reconstructed and, depending on experimental capabilities, pushed to rather low values, allowing a detailed characterisation of the hard-scale dependence of the suppression factor. Althought estimates of the latter are present in the literature for the specific process at hand~\cite{kope}, we take a conservative approach and avoid to introduce any suppression factor. We further assume factorisation to hold and adopt factorised expressions for the relevant cross sections. A preliminar set of, newly generated, diffractive parton distributions extracted from combined leading proton HERA data will be used for the calculation. In view of the expected factorisation breaking effects in hard, single-diffractive, measurements in hadronic collisions, the obtained values for the cross sections should be considered as upper bounds.

Given the explorative nature of the analysis, more intended as a feasibility study, 
theoretical predictions are calculated to leading order accuracy. 
We take into account, however, the virtual photon decay into leptons so that cross sections 
can be studied as a function of, measurable, final state leptons 
kinematics. This allows us to explore the phase space available for the process and to estimate
the impact of typical experimental cuts on the transverse momenta and rapidities of the leptons. 

From QCD analyses performed in DDIS and anticipating the results of the next section, we know that the colour-singlet exchanged in the $t$-channel is a gluon-enriched state.
Since gluonic contributions to Drell-Yan production starts to $\mathcal{O}(\alpha_s)$ in perturbation theory, 
an accurate estimation of the suppression factor will require the inclusion of higher order corrections. 
The impact of the latter and a detailed report on the extraction of diffractive parton distributions to NLO accuracy will be presented in a companion publication.

The paper is organised as follows. In Sec.~\ref{sec:2} we report in some details 
the extraction of diffractive PDFs from combined HERA leading proton data.
In Sec.~\ref{sec:3}, making use of such distributions, we present results for 
single-diffractive Drell-Yan production in $pp$ collisions at the LHC at $\sqrt{s}=13$ TeV.
In Sec.~\ref{Conclusions} we summarise our results. 

\section{FIT overview}
\label{sec:2}
\noindent
Diffractive DIS belongs to the Semi-Inclusive lepton-proton DIS process of the type 
\begin{equation}
\label{process}
l(k) \,+ \,p(P ) \,\rightarrow \,  l(k^{'} )\, + \,p(P')\, + \, X(p_X )\,,
\end{equation}
where, along with the outgoing lepton, an additional proton $p$ is detected in the final state.
In eq.~(\ref{process}) $X$ stands for the unobserved part of the hadronic final state and 
we indicate particles four-momenta  in parenthesis. 
In the $lp$ centre-of-mass system,  diffractive DIS events are then characterised
by outgoing protons with a large momentum fraction of the incident proton
and quite small values of the transverse momentum measured with respect to the collision axis, \textsl{i.e.} in the target fragmentation region of the incident proton.
The kinematic variables used to describe the DIS process are the conventional Lorentz invariants
\begin{equation}
Q^2=-q^2, \;\; x_B = \frac{Q^2}{2P \cdot q}, \;\; y=\frac{P \cdot q}{P \cdot k}\,,
\end{equation}
with $q=k-k'$. 
Final state protons are instead described by the
fractional momentum of the singlet exchange with respect to the proton momentum, $\xpom$, 
and the invariant momentum transfer $t$ at the proton vertex:
\begin{equation}
\xpom=\frac{q \cdot (P-P')}{P \cdot q} \,, \;\; t=(P-P')^2\,,
\end{equation}
where typical DDIS selection requires $\xpom \lesssim 0.1$ and $|t| \lesssim 1$ Ge$\mbox{V}^2$.
In the following we will use the scaled fractional momentum variable $\beta$ 
defined by $\beta=x_B/\xpom$. This is interpreted as the fractional momentum 
of interacting parton with respect to pomeron fractional momentum $\xpom$.
The data are often presented in terms of the reduced $lp$ cross section, $\sigma_r^{D(4)}$, which depends on the diffractive transverse and longitudinal structure functions $F_2^{D(4)}$ and $F_L^{D(4)}$, respectively. In the one-photon exchange approximation, it reads:
\begin{equation}
\label{sigmar}
\sigma_r^{D(4)}(\beta,Q^2,\xpom,t)=F_2^{D(4)}(\beta,Q^2,\xpom,t)-\frac{y^2}{1+(1-y)^2} F_L^{D(4)}(\beta,Q^2,\xpom,t)\,.
\end{equation}
According to the factorisation theorem~\cite{factorisation_soft,factorisation_coll}, structure functions 
appearing in eq.~(\ref{sigmar}), are factorised into perturbatively calculable short-distance cross sections and diffractive parton distributions
\begin{equation}
\label{hard_fact}
F_k^{D(4)}(\beta,Q^2,\xpom,t)=\sum_i \int_{\beta}^1
\frac{d\xi}{\xi} \; f_{i/p}^D(\beta,\mu_F^2;\xpom,t) \; C_{ki} 
\bigg( \frac{\beta}{\xi},\frac{Q^2}{\mu_F^2},\alpha_s(\mu_R^2)\bigg)
+\mathcal{O}\bigg( \frac{1}{Q^2}\bigg)\,.
\end{equation}
The index $i$ runs on the flavour of the interacting parton.
The hard-scattering coefficients $C_ {ki}$ ($k=2,L$) are pertubatively calculable 
as a power expansion in the strong coupling $\alpha_s$ and 
depend upon $\mu_F^2$ and $\mu_R^2$, the factorisation and renormalisation scales,
respectively. The $C_{ki}$ coefficient functions are the same as in fully inclusive DIS.
Diffractive PDFs $f_{i/p}^D(\beta,\mu_F^2,\xpom,t)$ appearing in eq.~(\ref{hard_fact}) are proton-to-proton fracture functions~\cite{trentadue_veneziano} 
in the very forward  kinematical limit and can be interpreted as 
the number density of interacting partons at a scale $\mu_F^2$ and fractional momentum $\beta$ conditional to the detection of a final state proton with fractional momentum $1-\xpom$ and invariant momentum transfer $t$.
The $t$-unintegrated diffractive PDFs appearing in eq.~(\ref{hard_fact}) obey 
standard DGLAP~\cite{DGLAP} evolution equations~\cite{extendedM}.
The same statement holds when they are integrated over $t$ in a limited range~\cite{newfracture}:
\begin{equation}
\label{intM}
f_{i/p}^D(\beta,Q^2,\xpom)=\int_{t_{min}}^{t_{max}} dt \, f_{i/p}^D(\beta,Q^2,\xpom,t)\,, \;\;\; t_{max} \ll Q^2\,.
\end{equation}
In this paper we analyse the combined H1 and ZEUS diffractive DIS cross sections measurements~\cite{H1ZEUS_combo_data} of the process in eq.~(\ref{process}) where leading protons are measured by dedicated forward spectrometers. The centre-of-mass energy for the $e^+p$ scattering is $\sqrt{s}=318$ GeV. 
This data set covers the phase space region $2.5 < Q^2 < 200$ Ge$\mbox{V}^2$ and 
$0.0018 < \beta < 0.816$ and it has widest coverage in the proton fractional energy loss, $0.00035 < \xpom < 0.09$, subdived in 10 bins in $\xpom$, with an average of 20 points per-$\xpom$ bin for 
a total of 192 points. At variance with all other DDIS cross sections measurements, the squared four-momentum transfer at the proton vertex, $t$, is integrated in the restricted range $0.09 < |t| < 0.55$ Ge$\mbox{V}^2$ 
in order to minimise systematic uncertainties originating from $t$-extrapolation of the various measurements outside their respective measured ranges.
The reduced cross sections in eq.~(\ref{sigmar}) are integrated over $t$ in such a range and 
the diffractive PDFs in eq.~(\ref{intM}) are defined accordingly.
For $\xpom<0.03$ the data set overlaps with high-statistics LRG data set
and for $0.03<\xpom<0.09$ it provides the best experimental information available on diffractive 
DIS cross sections. 
The combination procedure, in general, allows a reduction of the systematic uncertainties
via cross-calibration of the various measurements. The direct detection of the forward proton 
allows to avoid any systematics associated with the large rapidity gap selection. 
By definition, these data are free from the proton dissociative background which has been found to 
contribute around 23\% of the diffractive DIS cross sections based on LRG selection~\cite{H106LRG}. 
Therefore this set of data provides
the most precise knowledge about the absolute normalisation of diffractive DIS cross sections. 
These advantages however come at the price of 
increased uncertainties relative to LRG data given the reduced statistics of the sample.

\begin{table}[t]
\begin{center}
\begin{tabular}{c|c} \hline \hline
\hspace{0.5cm} Parameter\hspace{0.5cm} & \hspace{0.7cm} $p_i \pm \delta p_i$ \hspace{0.7cm}\\ \hline \hline
     $f_0$  &    -1.208    $\pm$   0.022  \\  
     $f_1$  &      48.2   $\pm$   11.9   \\ 
     $f_2$  &      1.42    $\pm$   0.13   \\ \hline
     $A_q$  &     0.0039   $\pm$   0.0007 \\       
     $B_q$  &    -0.237    $\pm$   0.026  \\ 
     $C_q$  &       0.5 \\
     $D_q$  &      22.6    $\pm$   2.8    \\     
     $E_q$  &     2.28     $\pm$   0.20   \\    \hline
     $A_g$  &     0.057     $\pm$   0.011  \\     
     $B_g$  &      0.41    $\pm$   0.13   \\ 
     $C_g$  &       0.5 \\ \hline
\end{tabular}
\hspace{1cm}
\begin{tabular}{c|c|c} \hline \hline
\hspace{0.5cm} $\xpom$ \hspace{0.5cm} & \hspace{0.5cm} $\chi^2$ \hspace{0.5cm} 
& \hspace{0.5cm} Fitted points \hspace{0.5cm}  \\ \hline \hline
  0.00035 &   4.44 &  4 \\
  0.0009 &   6.78 & 10 \\
  0.0025 &  21.36 & 16 \\
  0.0085 &  20.34 & 24 \\
  0.0160 &  20.70 & 26 \\
  0.0250 &  27.24 & 25 \\
  0.0350 &  13.85 & 24 \\ 
  0.0500 &  28.69 & 27 \\
  0.0750 &  13.10 & 26 \\
  0.0900 &  10.51 & 10 \\ \hline 
Total    &  167.0 & 192 \\ \hline 
\end{tabular}
\caption{\textit{Left: Best-fit parameters. Right: breakdown of $\chi^2$ contributions
in each $\xpom$ bin.}}
\label{best-fit-pars}
\end{center}
\end{table}

Diffractive parton distributions extracted form this data set will be used in the context of 
single hard diffraction in hadronic collisions in conjuction with ordinary parton 
distributions. In order to avoid any mismatch between inclusive and diffractive PDFs
we adopt leading order CTEQL1 parton distribution set~\cite{CTEQL1}
evolved in the zero-mass variable-flavour-number scheme (ZM-VFNS).
The evolution of diffractive PDFs is performed within the same scheme and to the same accuracy
by using \texttt{QCDNUM17}~\cite{QCDNUM17} program.
The QCD parameters are the ones quoted in Ref.~\cite{CTEQL1}. In particular
we set the charm and bottom masses to $m_c=1.3$ GeV
and $m_b=4.5$ GeV, respectively, and the strong coupling 
is evaluated at one loop setting $\alpha_s^{n_F=5}(M_Z^2)=0.130$.

\begin{figure}
\begin{center}
\includegraphics[scale=0.6]{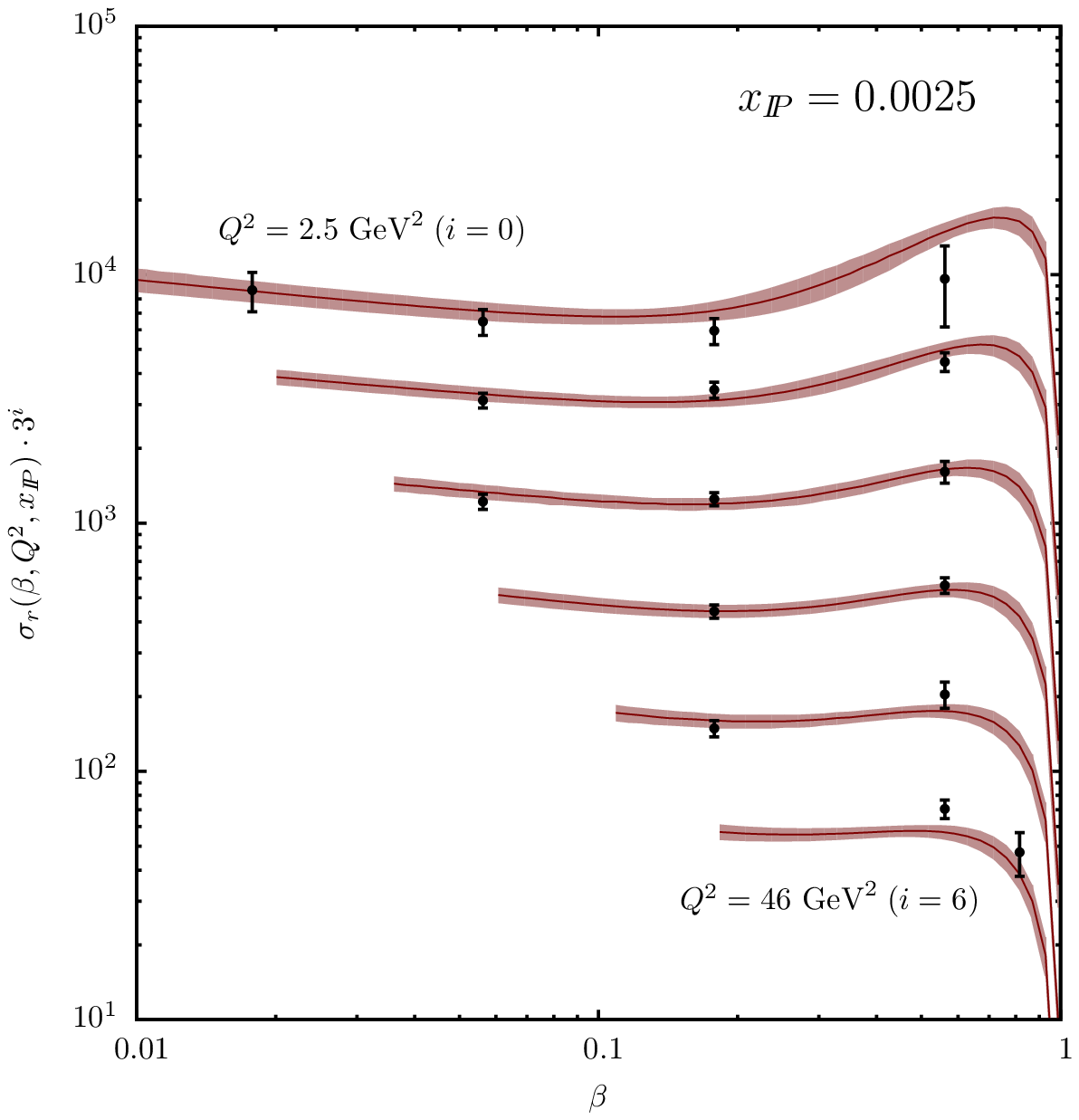}
\includegraphics[scale=0.6]{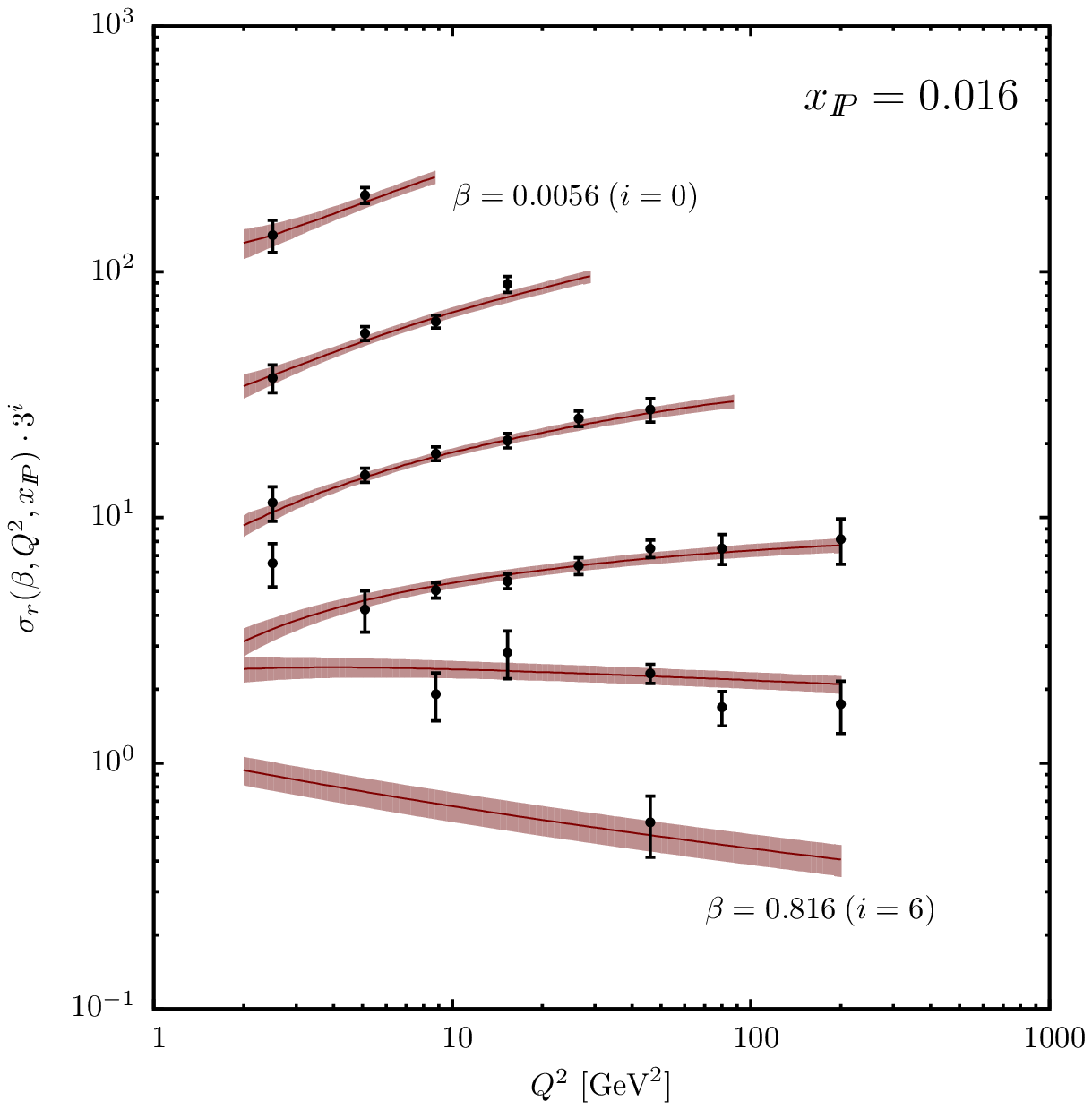}\\
\includegraphics[scale=0.6]{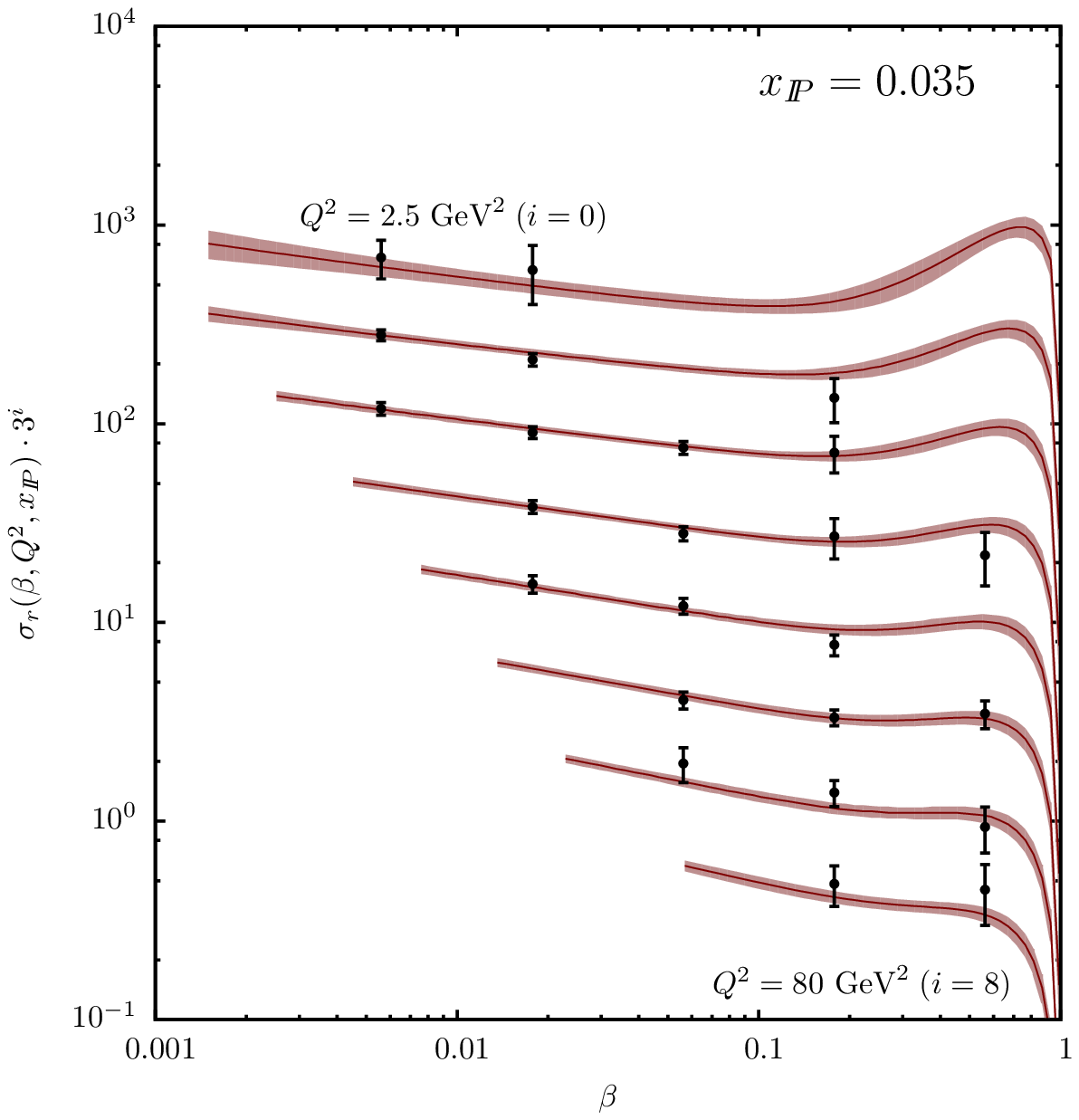}
\includegraphics[scale=0.6]{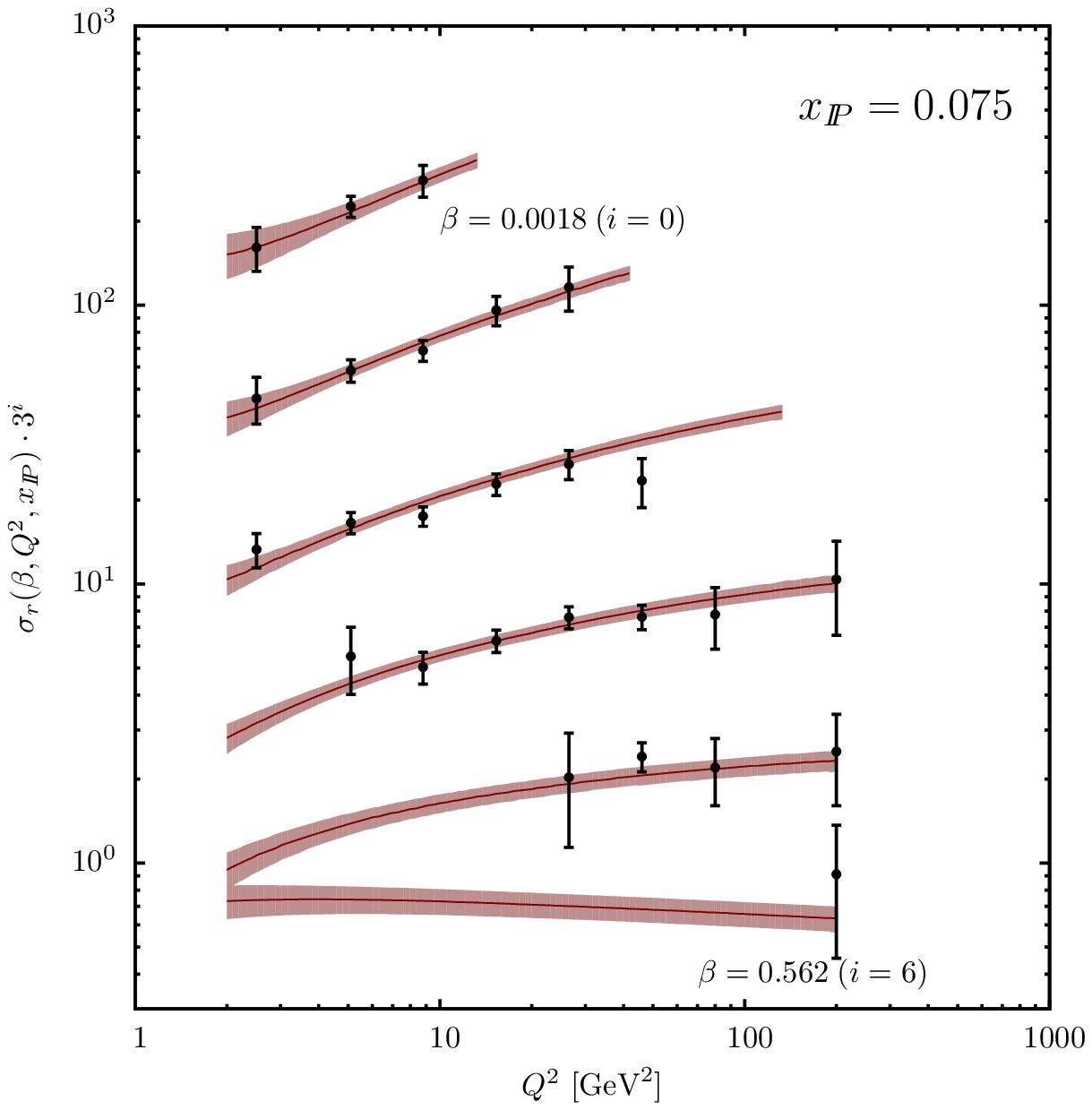}
\caption{\textsl{Best-fit results compared to combined H1-ZEUS data~\cite{H1ZEUS_combo_data}. 
The reduced cross section as a function as a function of $\beta$ or $Q^2$ is displayed in 
four representative bins of $\xpom$. Error bars are total uncertainties.
The band represents the propagation of experimental uncertainties according to  
the $\Delta_\chi^2=10$ criterion, as discussed in the text.}}
\label{Fig:sigma_r}
\end{center}
\end{figure}

In general factorisation theorem~\cite{factorisation_soft,factorisation_coll}
for diffractive DIS in the form of eq.~(\ref{hard_fact}) 
holds at fixed values of $\xpom$ and $t$ so that the parton content of the color-singlet exchange described by
$f_i^{D}$ is uniquely controlled by the kinematics of the outgoing proton.
Therefore, at least in principle, dPDFs may differ at different values of $\xpom$ and $t$. 
This idea has been successfully tested~\cite{myDPDF} in the analysis of LRG data from Ref.~\cite{H106LRG}.
In the present context, given the limited number and accuracy of the data points in each $\xpom$ bin, we use a simpler approach, namely a fully factorised $\beta-\xpom$ ansatz for the flavour-symmetric singlet and gluon diffractive parton (momentum) distributions defined at the initial scale $Q_0^2$:   
\begin{eqnarray}
\label{input}
\mathcal{F}(\xpom) &=&\xpom^{f_0} \; (1+f_1 \xpom^{f_2})\,, \nonumber\\
\beta \Sigma(\beta,Q_0^2,\xpom) &=& \mathcal{F}(\xpom) \; A_q \; \beta^{B_q} \; (1-\beta)^{C_q} (1+ D_q \beta^{E_q})\,, \\
\beta g(\beta,Q_0^2,\xpom) &=& \mathcal{F}(\xpom) A_g \; \beta^{B_g} \; (1-\beta)^{C_g}\,.\nonumber
\end{eqnarray}
The initial conditions in eq.~(\ref{input}) are characterised by a common 
flux factor $\mathcal{F}(\xpom)$ controlled by a single power at low $\xpom$.
An extra modulation, controlled by parameters $f_1$ and $f_2$,
is introduced to accomodate the data at larger values of $\xpom$. 
In order to guarantee the vanishing of the singlet distribution
on the endpoint, we fix the large-$\beta$ behaviour of the singlet 
by setting $C_q=0.5$ but additional freedom at intermediate values of $\beta$ is 
allowed leaving $D_q$ and $E_q$ parameters free in the minimisation. 
Since the gluon distributions is only indirectly fixed by the slope of the reduced cross section, the gluon parameters $B_g$ and $C_g$ are highly correlated and we decide to fix $C_g=0.5$
for a total of 9 free parameters. 
Such distributions, once evolved, are used to calculate the diffractive structure functions $F_{2,L}^D$
with the help of the \texttt{QCDNUM17} convolution engine and to reconstruct
the diffractive reduced cross sections in eq.~(\ref{sigmar}) which 
are then minimised against data~\cite{H1ZEUS_combo_data} with the help of the \texttt{MINUIT}~\cite{MINUIT} program. The choice of $Q_0^2$ is optimised performing a scan giving  
the best $\chi^2$ value for $Q_0^2=1.5 \, \mbox{GeV}^2$. 

\begin{figure}
\begin{center}
\includegraphics[scale=0.85]{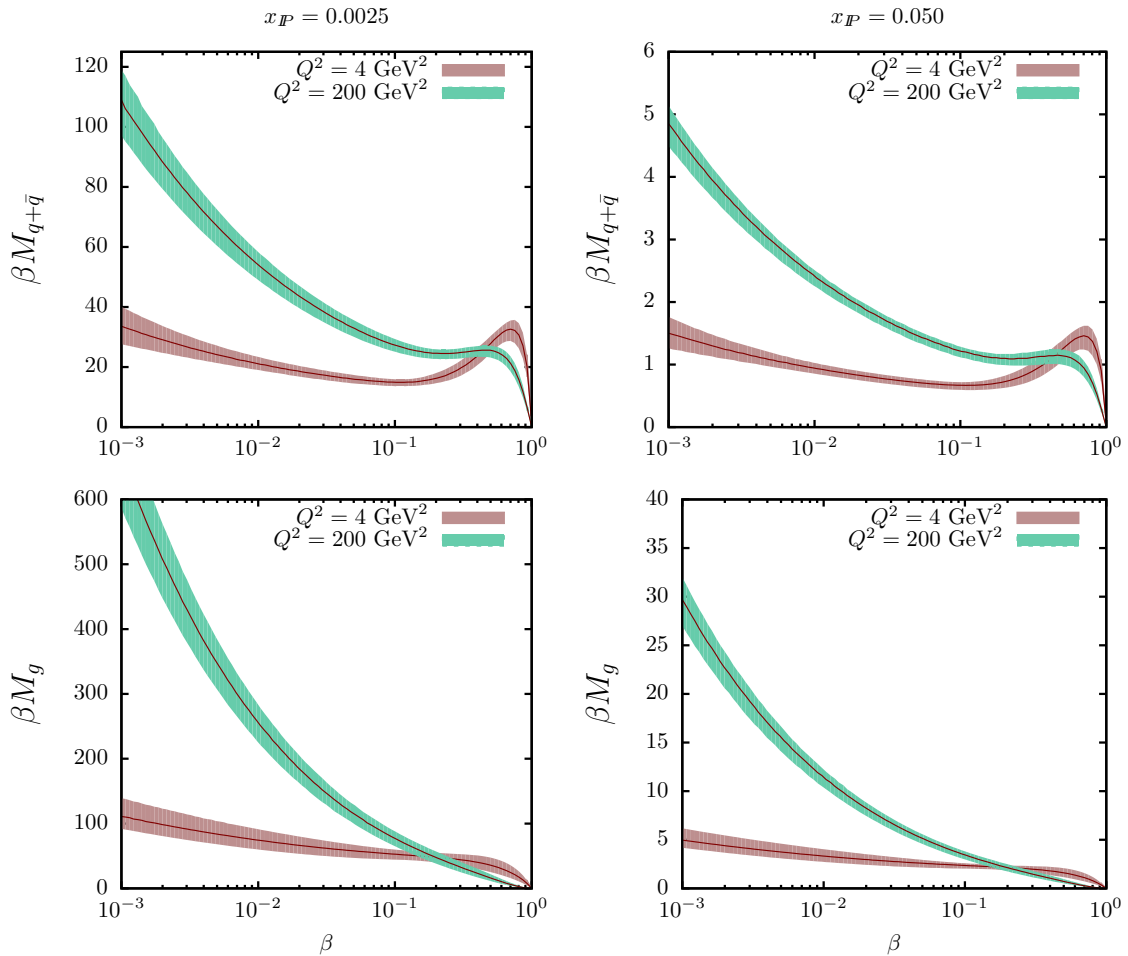}
\caption{\textsl{Diffractive parton distributions from best-fit evolved for two values of $Q^2$ and $x_\pom$ as a function of $\beta$. The band represents the propagation of experimental uncertainties according to  
the $\Delta_\chi^2=10$ criterion, as discussed in the text.}}
\label{Fig:xpdf_evo}
\end{center}
\end{figure}
\begin{figure}
\begin{center}
\includegraphics[scale=0.75]{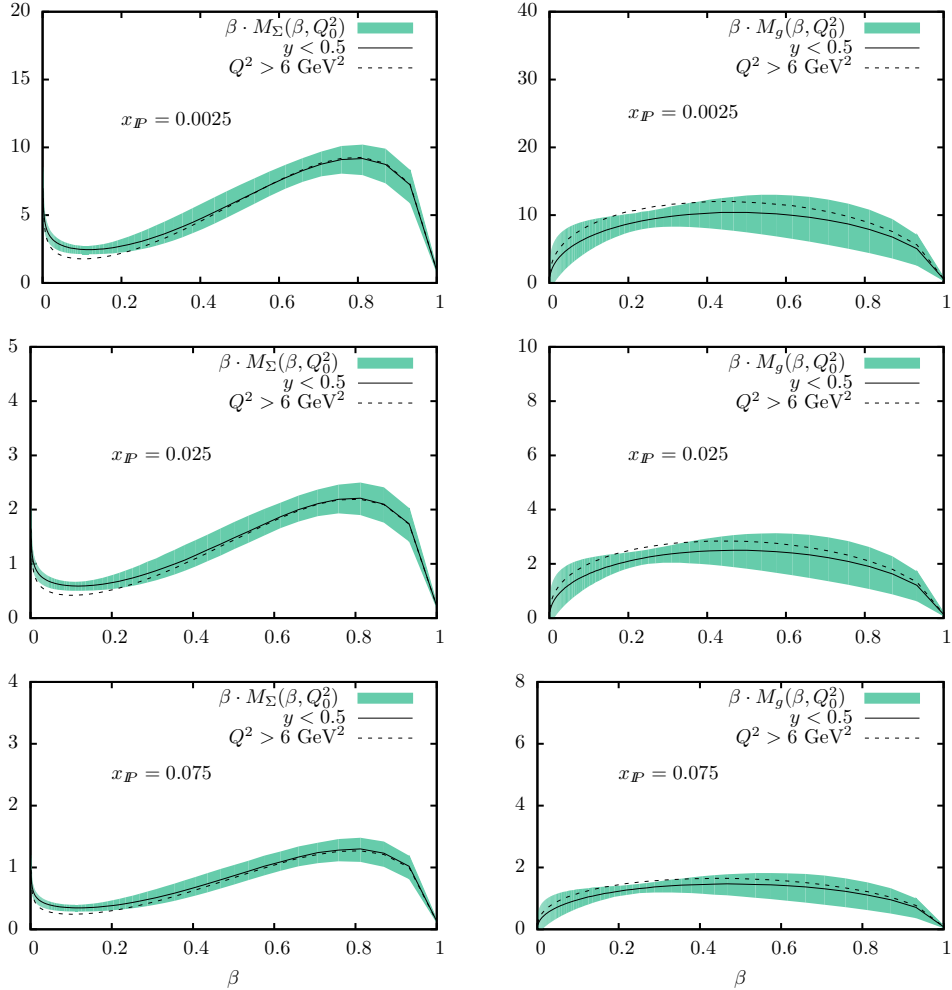}
\caption{\textsl{Diffractive singlet (left) and gluon (right) momentum distributions at $Q_0^2=1.5$ Ge$V^2$ for 
different values of $\xpom$. Best-fit distributions with uncertainties (the band corresponds to $\Delta\chi^2=10$) are compared with parametrisations returned by the fit with the cut $y<0.5$ (solid) or $Q^2>6$ Ge$V^2$ (dashed) imposed.}}
\label{Fig:xpdf_stability}
\end{center}
\end{figure}

By using total errors quoted in the experimental analysis 
and the standard $\chi^2$ definition, we obtain a total $\chi^2/d.o.f.$=0.913.
We report in Table~(\ref{best-fit-pars}) the best parameters and the breakdown of 
the contributions to $\chi^2$ function in each $\xpom$ bin. According to these numbers 
there is no misrepresentation of the data accross the $\xpom$ bins.
The comparison of the best fit results and the reduced cross sections is presented
in  Fig.~(\ref{Fig:sigma_r}) for four representative values of $\xpom$ as a
function of $Q^2$ or $\beta$.
We supplement the best-fit parametrisations with an additional set of parametrisations obtained 
according to the Hessian method outlined in Refs.~\cite{CTEQ_error,MRST_error} which allows to propagate experimental uncertainties to arbitrary observables. 
We note that our initial conditions assume a common $\beta$-shape for the diffractive PDF in all $\xpom$ bins. 
This theoretical hypothesis, in turn, determines an unrealistic precise determination of the diffractive PDFs
if associated with the standard $\Delta\chi^2=1$ criterion, often exceeding the precision of the data.
In order to correct for such an effect and to obtain a more conservative error estimate
we choose a tolerance criterion $\Delta\chi^2=10$ (one unit for $\xpom$ bin) and dPDFs 
alternative parametrisations are obtained with this choice.
We have checked by explicit evaluation that each parametrisation  
gives a consistent value for the $\chi^2$ function, $\chi_{\mbox{\tiny{best}}}^2+\Delta\chi^2$.
The error bands presented in the plots are obtained according to this criterion.

In Fig.~(\ref{Fig:xpdf_evo}) we present the singlet and gluon momentum distributions in two $\xpom$ bins at different scales. The singlet shows a bump in the large $\beta$ region ($\beta \gtrsim 0.5$) at the lower scale 
which is progressively washed away by evolution at higher scales. The rise of the gluon distribution at small
$\beta$ is accelerated by pQCD evolution and it is the dominant contribution for $\beta \lesssim 0.1$.      
 
We have further performed two consistency checks detailed below.
The first one concerns the diffractive longitudinal structure function which contributes starting from $\mathcal{O}(\alpha_s)$ and it is absent to the accuracy of the present calculation. Since its dominant contributions 
appear in the large-$y$ region, the fit has been repeated with the cut $y<0.5$ imposed.
The second one addresses the issue, reported in previous analyses 
~\cite{H106LRG,H107dijet,ZEUS09final,myDPDF}, of the inclusion in the fit of the lowest $Q^2$ points. 
For such a reason, the minimisation has been repeated by including only data points for which $Q^2>6$ Ge$\mbox{V}^2$. In both cases we observe a modest decrease in the $\chi^2/d.o.f$. 
However, as shown in Fig.~(\ref{Fig:xpdf_stability}), the resulting parametrisations are compatible, within uncertainties, with the ones obtained without imposing the cuts.
Given the substantial stability of the results against variation of the phase space
boundary of data included in the fit, we consider the ``no cut" scenario as our default choice and
use the corresponding best-fit parametrisations in the next Section.

\section{single-diffractive Drell-Yan production}
\label{sec:3}
The signature of hard diffraction in hadronic collisions is 
the presence of hard scattering process associated with the production of a leading proton.
Among many others, we consider here the simplest hard scattering process, namely 
the Drell-Yan pair production. Therefore we consider the reaction
\begin{equation}
\label{DDY}
p(P_1)+p(P_2) \rightarrow p(P)+\gamma^*(\rightarrow l^+(p_3)+l^-(p_4))+X\,.
\end{equation}
The invariant mass of the pair $q^2=(p_3+p_4)^2=Q^2$ is chosen to be large enough so that perturbative QCD can be applied.
In hadronic collisions, the Lorentz-invariant variable $z$ is used to characterise final state hadrons and is defined by
\begin{equation}
\label{z}
z=\frac{2P\cdot(P_1+P_2)}{s}\equiv \frac{2E_p^*}{\sqrt{s}}\equiv 1-\xpom\,.
\end{equation}
In the hadronic centre-of-mass frame, where the second identity holds, 
$z$ is just the observed proton energy, $E_p^*$, scaled down by the beam energy, $\sqrt{s}/2$.
Hard diffrative events are then characterised by low values of the invariant $\xpom$ and $t$, both 
in the same range of values as the one measured in DDIS.
 
Assuming factorisation to hold, one loop corrections to the process in eq.~(\ref{DDY}) have been reported in Refs~\cite{SIDYmy,SIDYmy2}, where a generalised procedure for the collinear factorisation is proposed. 
The latter is the same as the one proposed in particle production in the target fragmentation region in DIS~\cite{graudenz} and 
requires the introduction of fracture functions.  
However we stress that the ability to consistently subtract collinear singularities in such a semi-inclusive 
processes is a necessary but not a sufficient condition for factorisation to hold in hard diffractive processes in hadronic
collisions. The one-loop calculation mentioned above, infact, does take into account only the so-called active partons. 
It completely ignores multiple soft parton exchanges between active and spectators partons, 
whose effects should be accounted for in any proof of QCD factorisation (as done in the inclusive Drell-Yan case). 

In eq.~(\ref{DDY}), we assume that the proton with momentum $P_1$ is moving in the $+z$ direction
and the leading proton with momentum $P$ is produced quasi-collinearly to $P_1$ at large
and positive rapidities. At the cross section level, diffractive parton distributions 
for the proton with momentum $P_1$ will be used. The same process, of course, may occur also in the opposite emisphere
and, since the hadronic initial state is symmetric, will be not considered here. 

At the partonic level and to lowest order in the strong coupling, the process proceeds via the annihalition of a quark-antiquark pair into a massive virtual photon which subsequently decays into a opposite-sign lepton pair. To be definite 
we consider here the decay into muons:
\begin{equation}
q(p_1)+\bar{q}(p_2) \rightarrow \mu^+(p_3) + \mu^-(p_4)\,.
\end{equation}
Before discussing our results we found useful to sketch some details of the calculation.
The incoming parton momenta in the hadronic centre-of-mass-system are given by
\begin{equation}
p_1=x_1 \frac{\sqrt{s}}{2}(1,0_{\perp},1), \;\;\;\; p_2=x_2 \frac{\sqrt{s}}{2}(1,0_{\perp},-1)\,, 
\end{equation}
with $s=(P_1+P_2)^2$. We choose as final state variables the lepton rapidities $y_3$, $y_4$, and lepton transverse
momentum, $\bm{p_t}$. In terms of the latter, the four momenta of the leptons are given by
\begin{eqnarray}
p_3^{\mu}&=&(p_t \cosh y_3, \bm{p_t}, p_t \sinh y_3)\,,\\
p_4^{\mu}&=&(p_t \cosh y_4, -\bm{p_t}, p_t \sinh y_4)\,,\\
q^{\mu}&=&(M \cosh Y, \bm{0}, M \sinh Y)\,,
\end{eqnarray}
with $p_t=|\bm{p_t}|$ and $q=p_3+p_4$. The differential cross section, to leading order accuracy, 
involves appropriate products of diffractive and ordinary parton distributions functions. It reads
\begin{equation}
\label{DDYcs}
\frac{d\sigma^D}{dy_3 dy_4 dp_t d\xpom} =  \sum_q e_q^2 \frac{f_q^D(\beta,\xpom,\mu_F^2)}{\xpom} f_{\bar{q}}(x_2,\mu_F^2) \frac{2 p_t \hat{s}}{3s} \frac{2\pi\alpha_{em}^2}{\hat{s}^2}\frac{\hat{t}^2+\hat{u}^2}{\hat{s}^2} \,,
\end{equation}
where the sum runs over quark and antiquarks.
In actual calculations we have set the factorisation scale to $\mu_F=M_{\mu\mu}$. We further introduce the leptons rapidity sum, $Y$,  and difference $\bar{y}$: 
\begin{equation}
\label{Y}
Y=\frac{1}{2}(y_3+y_4)\,, \;\;\;\; \bar{y}=\frac{1}{2}(y_3-y_4).
\end{equation}
The former defines the rapidity of the virtual photon. 
The partonic Mandelstam invariants appearing in eq.~(\ref{DDYcs}) are then given by
\begin{eqnarray}
\hat{s}&=&p_t^2 (e^{\bar{y}}+e^{-\bar{y}})^2\equiv M_{\mu\mu}^2\,,\\
\hat{t}&=&(p_1-p_3)^2=-p_t^2 ( 1+ e^{-2\bar{y}})\,,\\
\hat{u}&=&(p_1-p_4)^2=-p_t^2 ( 1+ e^{2\bar{y}})\,.
\end{eqnarray}
In terms of these variables the momentum fractions are given by
\begin{eqnarray}
\beta&=&\frac{x_1}{\xpom}=\frac{p_t}{\xpom \sqrt{s}}(e^{y_3}+e^{y_4}) \equiv \frac{M_{\mu\mu}}{\xpom \sqrt{s}} e^Y \,,\\ 
x_2&=&\frac{p_t}{\sqrt{s}}(e^{-y_3}+e^{-y_4}) \equiv \frac{M_{\mu\mu}}{\sqrt{s}} e^{-Y}\,.
\end{eqnarray}
Since both momentum fractions can not exceed unity, the following bounds can be derived:
\begin{equation}
\label{DYrapidity_range}
\ln \sqrt{\tau} < Y < \ln \xpom -\ln \sqrt{\tau}\,,
\end{equation}
with $\tau=M_{\mu\mu}/\sqrt{s}$. Given the kinematic constraint $x_1 \leqslant x_{\!I\!P}$, 
the pair-rapidity spans an increasingly asymmetric range as $\xpom$ decreases. 
For $\xpom<\sqrt{\tau}$, the pair is entirely in the $Y<0$ rapidity range. 
Formally, the rapidity range for the inclusive Drell-Yan case 
is recovered simply setting $\xpom=1$ in eq.~(\ref{DYrapidity_range}).

In the present analysis we focus on diffractive processes tagged with 
dedicated instrumentation~\cite{LHC_rpot}.
We choose the proton fractional momentum loss to be in the range $10^{-4}<\xpom<10^{-1}$,
with maximal overlap with the range measured at HERA~\cite{H1ZEUS_combo_data}. 
Predictions presented in the following are integrated over the $t$-range of the 
data~\cite{H1ZEUS_combo_data} out of which dPDFs are estracted, \textsl{i.e.} $0.09<|t|<0.55$ Ge$\mbox{V}^2$.
We set the centre-of-mass energy of the $pp$ collisions to $\sqrt{s}=13$ TeV.
The invariant mass of the muon pair is required to be in the range $2<\mbox{M}_{\mu\mu}<20$ GeV, 
a range of virtualities in line with those measured at HERA. We assume that 
the $J/\Psi$ and $\Upsilon$  contributions, which both lie within this mass range, can be properly subtracted from the data sample. We require both muons to have rapidity  $|y^{\mu}|<2.45$ but we do not apply
cuts either on the muons transverse or three momenta.

\begin{figure}
\begin{center}
\includegraphics[scale=1.0]{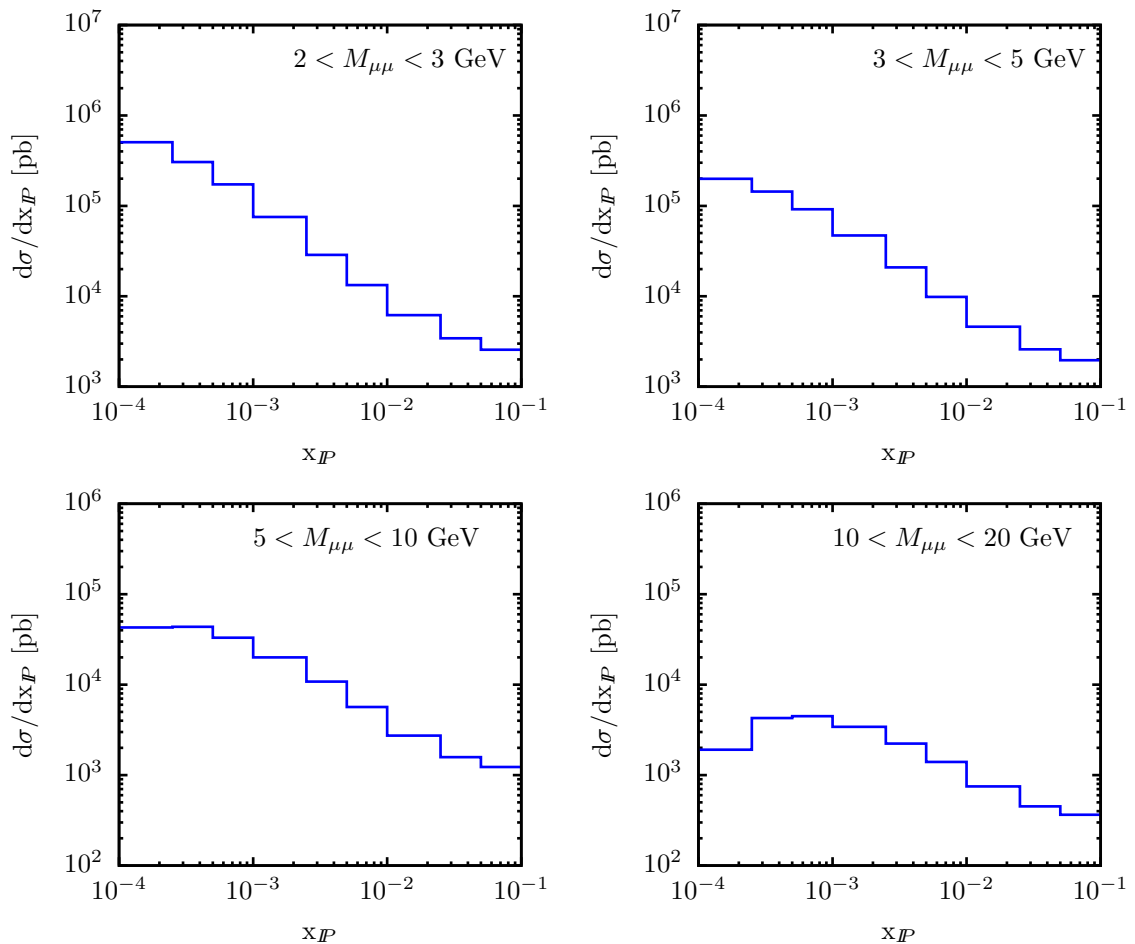}
\caption{Single diffractive DY production. DY $x_{I\!\!P}$-distribution in four mass ranges.}
\label{Fig:4}
\end{center}
\end{figure}

\begin{figure}
\begin{center}
\hspace{-0.3cm}
\includegraphics[scale=0.8]{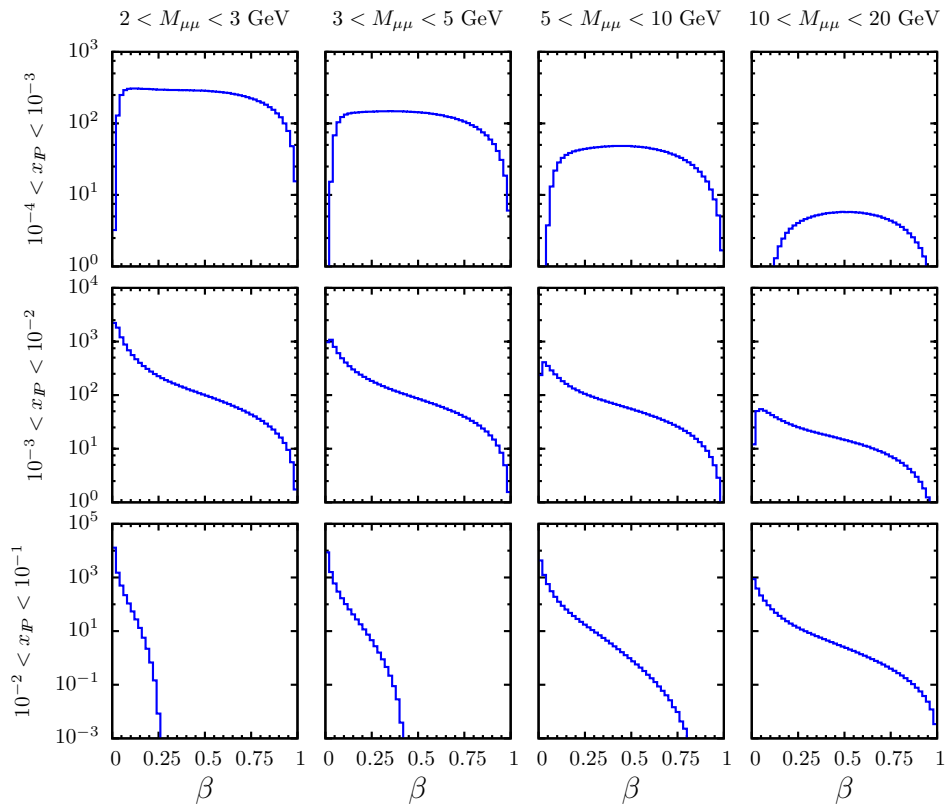}
\caption{\textsl{Single-diffractive Drell-Yan production. $\beta$-distribution in bins of $M_{\mu\mu}$ and $x_{I\!\!P}$}.}
\label{Fig:5}
\end{center}
\end{figure}

\begin{table}
\begin{tabular}{c|c} \hline \hline
Muon pair kinematics & $|y^{\mu}|<2.45$ \\
                     & $2<\mbox{M}_{\mu\mu}<20$ GeV \\
                     & No cuts on muon $p_t$ or $\bm{p}$\\ \hline
Proton kinematics  & $0.09<|t|<0.55$ Ge$\mbox{V}^2$ \\
                      & $10^{-4}<\xpom<10^{-1}$ \\  \hline  \hline
$\sigma^{SD,DY}$ & 1635 $\pm$ 60 (exp) $\phantom{}^{+650}_{-460}$ (scale) pb \\ \hline
\end{tabular}
\caption{\textsl{Outline of the muon pair and proton phase space regions and the corresponding fiducial cross section for single diffractive Drell-Yan pair production, $\sigma^{SD,DY}$}.}
\label{cut_and_cs}
\end{table}
The resulting fiducial cross sections for single-diffractive Drell-Yan pair production is  
reported in Tab.~(\ref{cut_and_cs}). In the case that proton spectrometers are installed on both side of the interaction point, such a result for the fiducial cross section should be doubled. As already anticipated, 
the quoted result does not include any rapidity gap suppression factor and 
predictions refer to virtual photon decay in the muon channel.
The first error represents the propagation of experimental uncertainties as obtained 
in the diffractive PDF fit. The second one, of theoretical nature, is obtained 
varying the factorisation scale $\mu_F^2$ appearing in both distributions in eq.~(\ref{DDYcs}) 
in the range $1/2 M_{\mu\mu}^2 < \mu_F^2 < 2 M_{\mu\mu}^2$. 
In this regime of relatively low $Q^2$ values where diffractive and inclusive parton 
distributions evolve faster, we find that the dominant error source, of theoretical nature, is associated with missing higher order corrections. The latter are known to high accuracy for a number of distributions
and will be considered in a separate publication. In the present contest, predictions
can be stabilised against factorisation scale variation by considering appropriate ratios 
of diffractive over inclusive cross sections.
This issue will be discussed in some details at the end of this Section. 
\begin{figure}
\begin{center}
\includegraphics[scale=1.0]{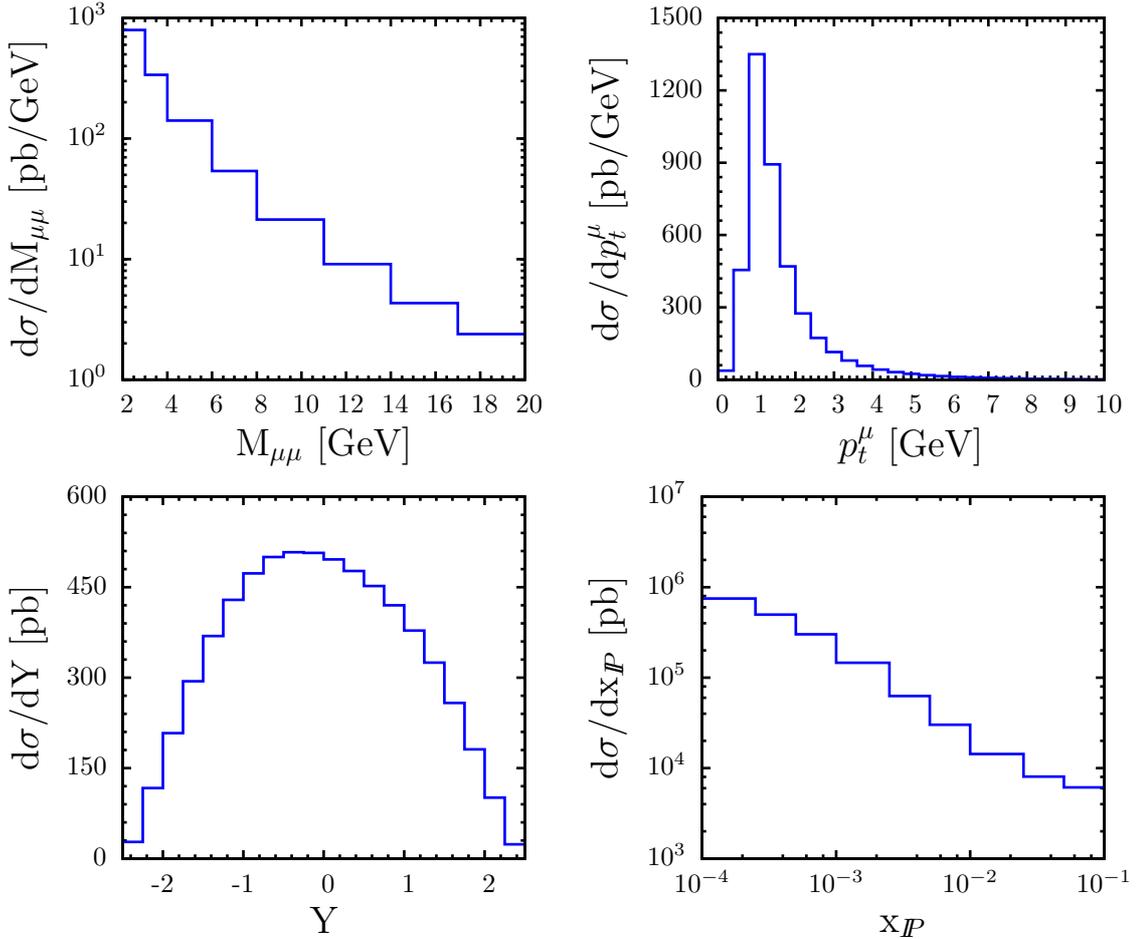}
\caption{\textsl{Single-diffractive Drell-Yan production. Top left: invariant mass distribution.
Top right: transverse momentum distribution of final state muons.
Bottom left:  muon pair rapidity distribution. Bottom right: $\xpom$ distribution.}}
\label{Fig:1}
\end{center}
\end{figure}
We begin our overview of predictions showing in Fig.~(\ref{Fig:1}) single-differential cross sections
in the fiducial phase space region specified in Tab.~(\ref{cut_and_cs}).
The pair invariant-mass distribution is shown in the top left panel
and rapidly falls as an inverse power of $\hat{s}=M_{\mu\mu}^2$ 
typical of annihilation processes into massive states, 
as it can be read out from eq.~(\ref{DDYcs}).
In the top right panel the muon transverse momentutm distribution is presented. 
Its kinematically allowed range extend up to $p_t=M_{\mu\mu}^{max}/2$.
Given the fast falling nature of the $M_{\mu\mu}$-distribution,
dominated by low values of the invariant, the muon transverse momentum distribution shows 
a maximum (the Jacobian peak) at $p_t=M_{\mu\mu}^{min}/2$.
The muon pair rapidity distribution, presented in the bottom left panel, is
slightly asymmetric and indicates a preference for the virtual 
photon to populate the negative rapidity emisphere (the one containing the dissociated proton, in the chosen reference frame).
\begin{figure}
\begin{center}
\includegraphics[scale=0.8]{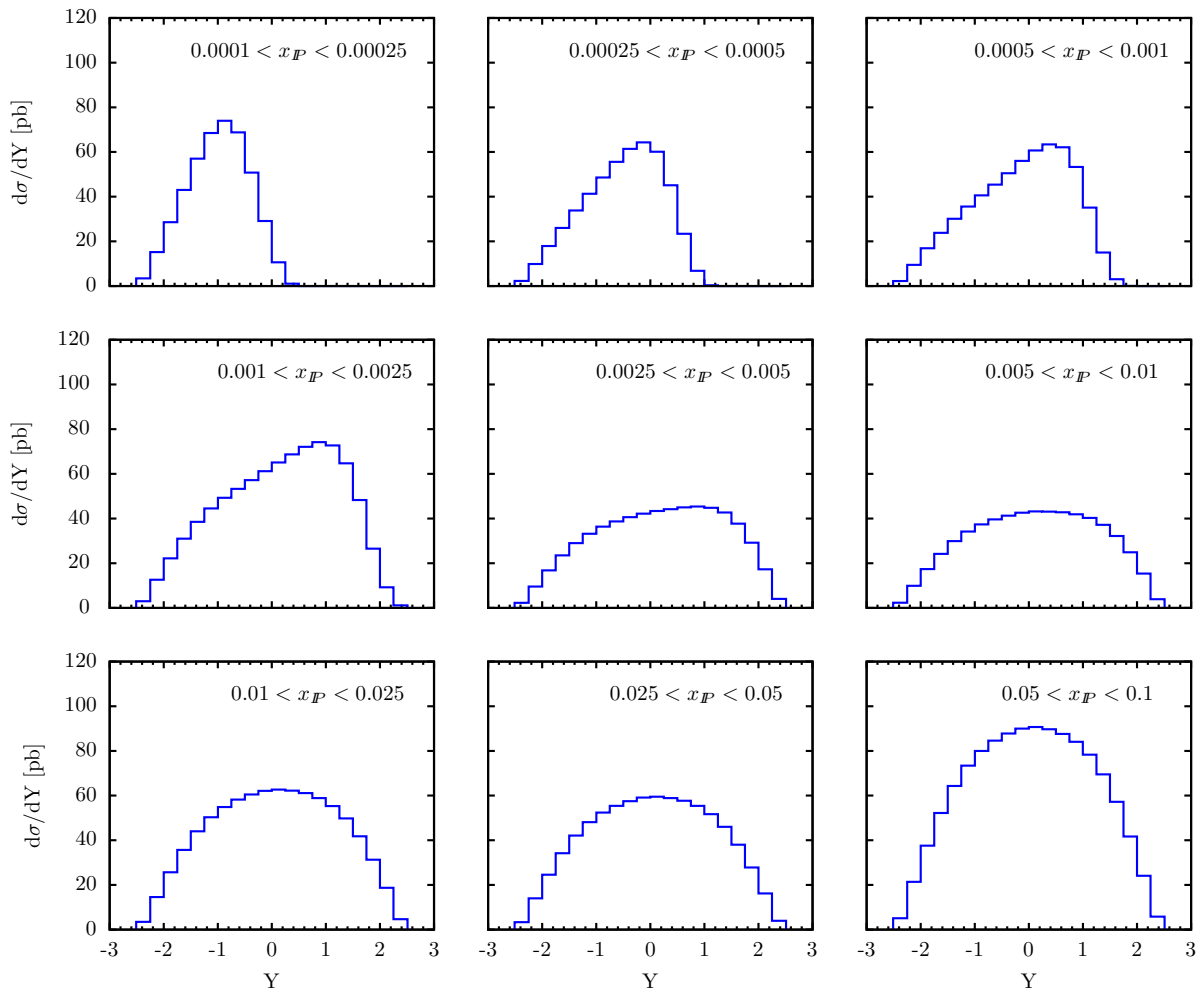}
\caption{\textsl{Moun-pair rapidity distribution in bins of $x_{I\!\!P}$ integrated over the fiducial range  
$2<M_{\mu\mu}<20$ GeV.}}
\label{Fig:2}
\end{center}
\end{figure}
We note that, despite phase space limitations introduced by eq.~(\ref{DYrapidity_range})
and the difference between diffractive and ordinary parton distributions, 
the muon-pair populates the available rapidity range, as defined by the muon rapidity cuts
and by eq.~(\ref{Y}). In the bottom right panel we present the $\xpom$ distribution. 
In general, it is well known that such distribution behaves approximately as an inverse 
power of $\xpom$ at small $\xpom$. In the present case, the flattening of the distribution
at small $\xpom$ can be ascribed to the shrinkage of phase space for the production 
of massive pair, since the maximum partonic centre-of-mass energy is reduced to $\sqrt{\xpom s}$.
The kinematic of the scattered proton induces peculiar features on muon pair production
whose effects are presented in Fig.~(\ref{Fig:2}) in terms of the muon pair rapidity, Y, 
in various bins of $\xpom$.
The distributions is strongly asymmetric at the lowest values of $\xpom$,
where the muon pair populates the negative rapidity range (dissociated proton direction) due to the kinematic constrain $x_1 < \xpom$. 
\begin{figure}
\begin{center}
\includegraphics[scale=1.0]{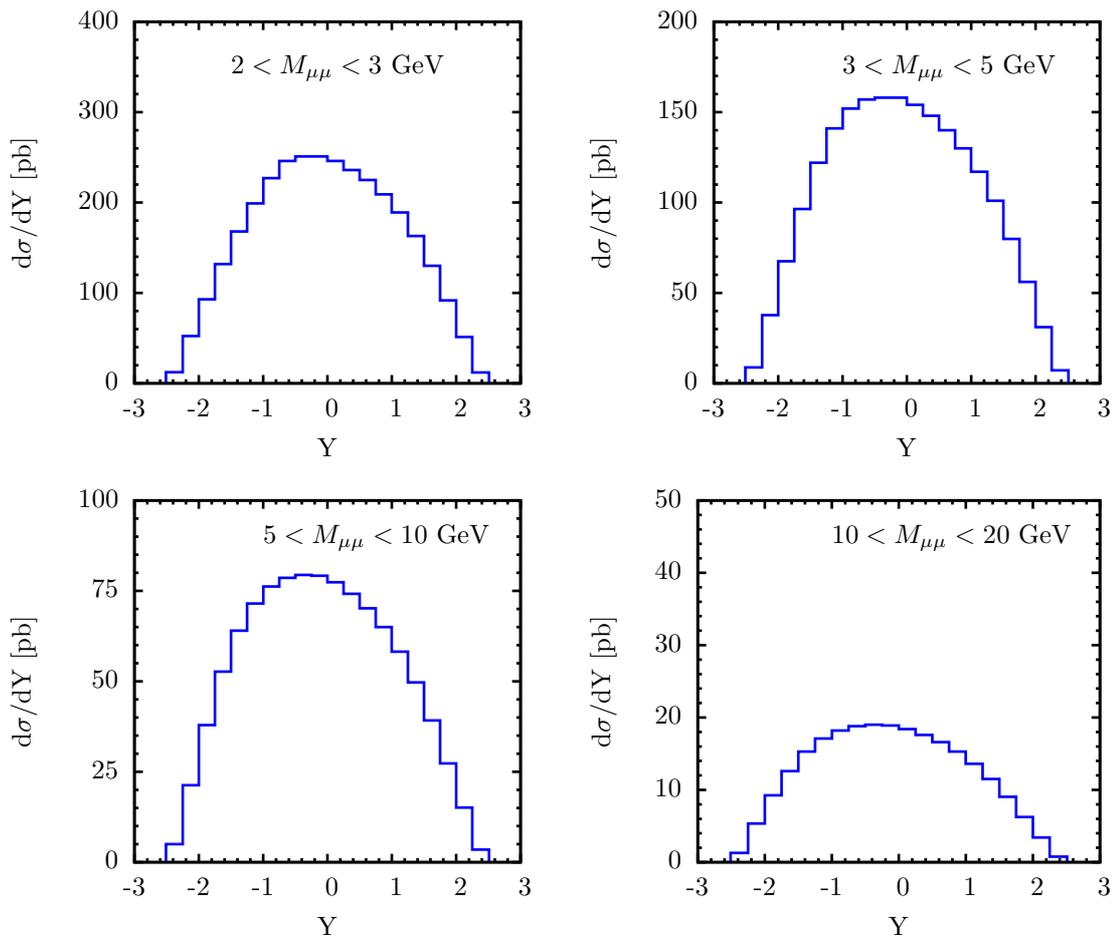}
\caption{\textsl{Single diffractive DY production. Muon-pair rapidity distribution in bins of $M_{\mu\mu}$}.}
\label{Fig:3}
\end{center}
\end{figure}
In the intermediate $\xpom$ range the pair starts to populate the positive emisphere (diffractive proton direction) with a tendency to show a maximum in this range. 
At even higher values of $\xpom$, the available centre-of-mass for the reaction increases
and the distribution progressively turns into a symmetric one.
Given the relatively light masses produced, this regime is sensitive to parton distributions 
evaluated at relatively small values of $\beta$ and $x_2$, the symmetry of the rapidity distribution
indicates that the shapes of the sea component both of diffractive and ordinary distributions are similar, being both driven by QCD evolution. 
This complicated pattern is further illustrated in Fig.~(\ref{Fig:3}) where 
the single differential cross section as a function of $Y$ is shown in four different 
ranges of the pair invariant mass and integrated over $\xpom$. In all mass bins, 
the distributions show a maximum in the negative rapidity range,
a signal that the interacting parton from the dissociated proton carries,
on average, slightly more momentum with respect to the one originating from 
the scattered proton.
In Fig.~(\ref{Fig:4}) we present single differential distributions as a function of $\xpom$
in four different invariant mass ranges. As the invariant mass increases, we observe
a progressive flattening of the distributions at small $\xpom$. This effect is due 
to the phase space reduction induced by the constrain $M_{\mu\mu}^2=\beta \xpom x_2 s$, 
which at low $\xpom$ disfavours the production of increasingly massive pair.
In Fig.~(\ref{Fig:5}) we present single differential cross section as a function of $\beta$, 
the fractional momentum of the interacting parton with respect to the one of the colour singlet exchanged in the $t$-channel, integrated in various bins of $M_{\mu\mu}$ and $\xpom$. 
Such distributions offer an insight to the sensitivity of the 
cross section to diffractive parton distributions, modulo kinematics effects.
In the lowest $\xpom$ bin the distributions span all the allowed $\beta$ range  
and progressively shrinks at large $\beta$ as $\xpom$ increases, a natural consequence 
of momentum conservation.  
As already shown in Fig.~(\ref{Fig:2}) and Fig.~(\ref{Fig:5}), the distributions in the pair rapidity $Y$ are asymmetric around $Y=0$. The asymmetry decreases both as the mass of the pair increases and as $\xpom$ increases. Such an effect is absent in the inclusive 
Drell-Yan case initiated by a symmetric initial state. 
\begin{figure}
\begin{center}
\includegraphics[scale=0.7]{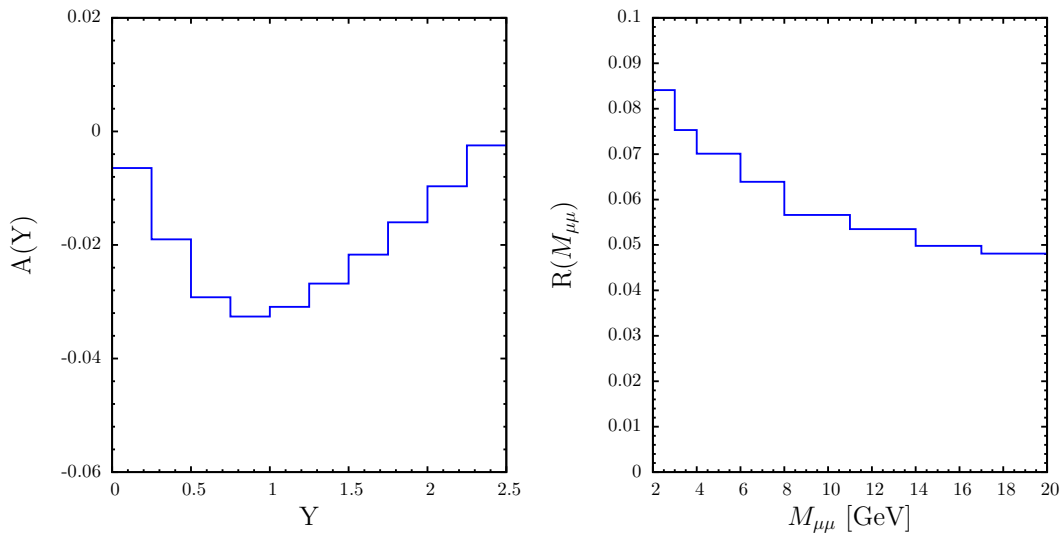}
\caption{\textsl{Left panel: asymmetry of the pair-rapidity distribution. Right panel: ratio of diffractive over inclusive Drell-Yan cross sections as a function of $Q^2$.}}
\label{Fig:7}
\end{center}
\end{figure}
This effect is better appreciated considering the asymmetry $A$ defined by
\begin{equation}
\label{asy}
A(Y)=\frac{d\sigma(Y)-d\sigma(-Y)}{d\sigma(Y)+d\sigma(-Y)}\,.
\end{equation} 
and shown in the left panel of Fig.~(\ref{Fig:7}). 
The asymmetry, integrated over all masses and proton energy loss, the asymmetry reaches its maximum 3\% at  $Y\simeq 1$ and is always negative, implying that the muon-pair 
is produced mostly in the emisphere opposite to the one containing the scattered proton.
The asymmetry, being normalised to the integrated single diffractive cross section, is not affected by uncertainties due to the rapidity gap survival factor and it is sensitive to the shape of diffractive PDFs.
Depending on the accuracy of the data, this predicted behaviour, absent in the inclusive case,  could be expolited to correlate the forward proton detection with the central Drell-Yan production.  As discussed at the beginning of the Section, predictions are affected by 
large theoretical errors associated with scale variations. 
Such uncertainties can be conveniently reduced by considering 
the  ratio $R$ of diffractive to inclusive cross sections
\begin{equation}
\label{ratio}
R=\frac{\sigma (p p \rightarrow p XY)}{\sigma(p p\rightarrow XY)}\,,
\end{equation}
which also offers the advantage to reduce experimental systematics errors.
In eq.~(\ref{ratio}) $Y$ stands for the selected hard scattering process (DY this case)
and $X$ for the unobserved part of the final state. 
At Tevatron the ratio $R$ has been measured in a variey of final state~\cite{diff_bbar,diff_dijet,diff_WZ} and it shows a quite stable behaviour with a value close to 1\%. 
For the single-diffractive Drell-Yan production  in $pp$ collisions at $\sqrt{s}=13$ TeV, 
the ratio $R$ is presented in the right panel of Fig.~(\ref{Fig:7}).
Given our leading order estimate of the inclusive Drell-Yan cross section, 
$R$ varies between 6\% and 8\% and decreases mildly as a function of the invariant mass of the pair, $M_{\mu\mu}$. This prediction however does not take into account the RGS suppression factor. With this respect it would be interesting to check whether 
the data follow at least the shape of the ratio as a function of $M_{\mu\mu}$.

\section{Conclusions}
\label{Conclusions}
In this paper we have considered the single-diffractive production of low-mass Drell-Yan 
pair in $pp$ collisions at the LHC at $\sqrt{s}=13$ TeV. Predictions are based on a fully factorised
approach for the cross section which makes use of a set of diffractive 
parton distributions obtained from a QCD fit to combined leading proton DIS data from HERA.
A number of distributions are presented both in terms of Drell-Yan 
pair and scattered proton variables. Examples of asymmetries and ratio are constructed
in order to minimise theoretical and experimental uncertanties. 
In view of the foreseen measurements of this type of process at the LHC Run-II, 
these results constitute a baseline for the characterisation of the expected factorisation breaking effects.  
\section*{Acknowledgements}
\noindent
This work is supported in part through the project 
``Hadron Physics at the LHC: looking for signatures of
multiple parton interactions and quark gluon plasma formation (Gossip project)",
funded by the ``Fondo ricerca di base di Ateneo" of the Perugia University.
We warmly thank Marta Ruspa for numerous discussions
on the LHC measurements and for providing us details of the analysis presented in Ref.~\cite{H1ZEUS_combo_data}.
We also thank Sergio Scopetta for reading the manuscript before submission.

\end{document}